\documentclass[twocolumn,english,aps,floatfix,superscriptaddress]{revtex4}
\usepackage[T1]{fontenc}
\usepackage[latin9]{inputenc}
\usepackage{amsmath}
\usepackage{graphicx}
\usepackage{esint}

\makeatletter
\@ifundefined{textcolor}{}
{%
 \definecolor{BLACK}{gray}{0}
 \definecolor{WHITE}{gray}{1}
 \definecolor{RED}{rgb}{1,0,0}
 \definecolor{GREEN}{rgb}{0,1,0}
 \definecolor{BLUE}{rgb}{0,0,1}
 \definecolor{CYAN}{cmyk}{1,0,0,0}
 \definecolor{MAGENTA}{cmyk}{0,1,0,0}
 \definecolor{YELLOW}{cmyk}{0,0,1,0}
}

\usepackage{graphics}

\@ifundefined{definecolor}{\@ifundefined{definecolor}
 {\usepackage{color}}{}
}{}

\usepackage{bm}

\usepackage{babel}

\usepackage{babel}

\usepackage{babel}

\usepackage{babel}

\@ifundefined{definecolor}{\@ifundefined{definecolor}
 {\@ifundefined{definecolor}
 {\@ifundefined{definecolor}
 {
}{}
}{}
}{}
}{}

\usepackage{babel}

\usepackage{babel}

\makeatother

\usepackage{babel}
\begin{document}

\title{Two-band superconductivity in doped SrTiO$_{3}$ films and interfaces}

\author{R. M. Fernandes}

\affiliation{Department of Physics, Columbia University, New York, New York 10027,
USA }

\affiliation{Theoretical Division, Los Alamos National Laboratory, Los Alamos,
NM, 87545, USA}

\author{J. T. Haraldsen}

\affiliation{Theoretical Division, Los Alamos National Laboratory, Los Alamos,
NM, 87545, USA}

\affiliation{Center for Integrated Nanotechnologies, Los Alamos National Laboratory,
Los Alamos, NM 87545, USA}

\author{P. Wölfle}

\affiliation{Institute for Condensed Matter Theory and Institute for Nanotechnology,
Karlsruhe Institute of Technology, D-76128 Karlsruhe, Germany}

\author{A. V. Balatsky}

\affiliation{Theoretical Division, Los Alamos National Laboratory, Los Alamos,
NM, 87545, USA}

\affiliation{Center for Integrated Nanotechnologies, Los Alamos National Laboratory,
Los Alamos, NM 87545, USA}
\begin{abstract}
We investigate the possibility of multi-band superconductivity in
SrTiO$_{3}$ films and interfaces using a two-dimensional two-band
model. In the undoped compound, one of the bands is occupied whereas
the other is empty. As the chemical potential shifts due to doping
by negative charge carriers or application of an electric field, the
second band becomes occupied, giving rise to a strong enhancement
of the transition temperature and a sharp feature in the gap functions,
which is manifested in the local density of states spectrum. By comparing
our results with tunneling experiments in Nb-doped SrTiO$_{3}$, we
find that intra-band pairing dominates over inter-band pairing, unlike
other known multi-band superconductors. Given the similarities with
the value of the transition temperature and with the band structure
of LaAlO$_{3}$/SrTiO$_{3}$ heterostructures, we speculate that the
superconductivity observed in SrTiO$_{3}$ interfaces may be similar
in nature to that of bulk SrTiO$_{3}$, involving multiple bands with
distinct electronic occupations. 
\end{abstract}
\maketitle

\section{Introduction}

The nature of multi-band superconductivity has been the subject of
intense experimental\cite{Bednorz80,lava:02} and theoretical\cite{suhl:59,scho:77,iski:07,babaev:12,geye:10}
debate for decades. In the mid-1960s, specific heat measurements found
evidence of multiple superconducting gaps in some elemental superconductors,
such as Nb and V. More recently, multi-band superconductivity has
been found in compounds displaying relatively high transition temperatures
$T_{c}$: the magnesium diborides, with $T_{c}\approx39$K\cite{Mazin11},
and the iron pnictides, with $T_{c}$ up to $56$K\cite{kami:08,chen:08,ren:08,cheng:08,si:08,rott:08}.
While the superconducting states in these two classes of materials
have significantly different microscopic structures, many of their
thermodynamic properties display common signatures associated with
the existence of multiple gaps. Such features encompass the temperature
dependence of the specific heat and of the penetration depth, as well
as the multiple-peak structure in tunneling spectroscopy.

\begin{figure}
\includegraphics[width=3in]{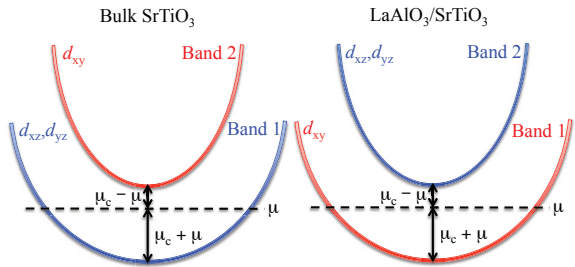} \caption{(Color Online) A general illustration of the two band model as applied
to bulk STO (left panel) and LAO/STO (right panel). For STO, the $d_{xz/yz}$
(band 1)crosses the Fermi level and the bottom of the $d_{xy}$ (band
2) is located above the Fermi level by $\mu_{c}-\mu$. In the case
of LAO/STO interface, orbital reconstruction switches the position
of the bands.}

\label{2band} 
\end{figure}

Some of these characteristic multi-gap features have been observed
in a lesser investigated material, electron-doped SrTiO$_{3}$ (STO).
Indeed, oxygen-deficient STO possesses a superconducting state below
$T_{c}\approx0.3-0.4$ K\cite{koon:67,Bednorz80}. Remarkably, tunneling
measurements by Binnig $et~al.$ \cite{Bednorz80} in Nb-doped STO
in the early 1980s found two peaks in the local density of states
beyond a certain electronic density, providing strong evidence for
multi-band superconductivity. More interestingly, the appearance of
a second superconducting gap is correlated with a sudden increase
in $T_{c}$, which can reach values up to $0.7$K. Band structure
calculations on Nb-doped STO also finds that near the electronic concentration
where $T_{c}$ is maximum, an additional electron-like band crosses
the Fermi level \cite{Mattheiss72,Mazin11}.

Besides bulk STO, the recently discovered heterostructures of LaAlO$_{3}$/SrTiO$_{3}$
(LAO/STO) also display superconductivity up to $T_{c}\sim0.2$ K\cite{reyr:07,hebe:09},
which can be enhanced to $0.3-0.4$ K by the application of an electric
field \cite{cavi:08,bell:09}. This is surprisingly close to the aforementioned
$T_{c}$ value in bulk STO. Recent experiments have also found strong
evidence that the maximum $T_{c}$ in the heterostructures is achieved
once an extra electron band becomes occupied\cite{Ilani11}. In view
of the early experiments by Binnig $et~al.$ \cite{Bednorz80}, it
is tempting to trace an analogy between superconductivity in bulk
STO and in the LAO/STO heterostructures. Thus, the understanding of
the multi-band superconducting state in electron-doped STO may be
relevant for the ongoing debate on the nature of the superconducting
LAO/STO heterostructures.

In both systems - bulk STO and LAO/STO interfaces - the origin and
mechanism of the SC state is still an open issue. Most of the studies,
however, have employed a single-band approach \cite{Mazin11,huij:09,gari:09,zubk:11,bert:11}.
In this paper, we do not discuss the origin of the SC state. We instead
use a mean-field phenomenological two-dimensional model to focus on
the consequences and signatures of multi-band superconductivity in
these systems for varying electronic concentration. Therefore, our
approach is most relevant for thin films of doped STO and interfaces
of STO with other systems that may provide negative charge carriers,
such as LaAlO$_{3}$ and interfacial oxide gel\cite{ueno:08}.

In our model, motivated by first-principle calculations \cite{Mattheiss72,Mazin11},
the $d_{xz,yz}$ and $d_{xy}$ orbitals form two electron bands such
that, in the weakly doped compound, only one of them crosses the Fermi
level (see Fig. \ref{2band}). As the system is doped with negative
charge carriers and the chemical potential $\mu$ shifts upwards,
the second band crosses the Fermi level, causing important changes
in the SC properties of the material. We calculate $T_{c}$, the gap
functions, and the local density of states (LDOS) for different electronic
concentrations, and compare our results with the tunneling experiments
by Binnig $et~al.$ in Nb-doped STO \cite{Bednorz80}. Such a comparison
demonstrates that an attractive \emph{intra-band} (\emph{intra-orbital})
pairing interaction dominates over the \emph{inter-band} (\emph{inter-orbital})
pairing, in contrast to other recently found multi-band superconductors,
such as the iron pnictides.

Although small, the inter-band coupling causes a sharp increase of
$T_{c}$ when the second band is occupied at $\mu=\mu_{c}$, and also
enhances both gap functions, with the gap in the initially unoccupied
band increasing faster than the gap in the initially occupied band.
As a result, although for the undoped system these gaps have very
different orders of magnitude, beyond $\mu=\mu_{c}$ they achieve
comparable values, being manifested in the partial gaps as functions
of $\mu$ and $T$ (Fig. \ref{Egap-3DLDOS}(a)). Our results are independent
of the character of the inter-band interaction: for an attractive
(repulsive) interaction, the gaps on two bands have the same (opposite)
signs, but the thermodynamic properties discussed here remain the
same. We also discuss possible connections with STO interfaces, proposing
tunneling experiments on both the LAO/STO heterostructures and STO
thin films to shed light on the superconducting state of these heterostructures.
Finally, we discuss the importance of the dimensionality, arguing
that most of the sharp features observed at $\mu=\mu_{c}$ in the
two-dimensional model may be smoothed out by three-dimensional band
dispersions.

Our paper is organized as follows: in Section \ref{sec_model} we
formulate and solve our two-dimensional two-band model, discussing
how $T_{c}$ and the gap functions change as the chemical potential
is varied. In Section \ref{sec_tunneling}, we calculate the LDOS
corresponding to the two-band model, comparing our results to the
experiments of Binnig \emph{et al \cite{Bednorz80}. }The effects
of three-dimensionality are discussed in Section \ref{sec_three_dimensional},
and our concluding remarks are presented in Section \ref{sec_conclusions}.

\section{Two-dimensional two-band model\label{sec_model}}

\subsection{Formulation of the model}

First-principle calculations \cite{Mattheiss72,Mazin11} reveal that
the band structure of bulk STO near the Fermi level is composed of
the Ti $t_{2g}$ orbitals $d_{xz}$, $d_{yz}$, and $d_{xy}$. Due
to the spin-orbit coupling, there is a splitting of $2\mu_{c}\sim20-30$
meV between the $d_{xz/yz}$ and $d_{xy}$ orbitals, with the latter
having a larger onsite energy than the former. There is also another
splitting between the $d_{xz}$ and $d_{yz}$ orbitals, which is however
one order of magnitude smaller than $\mu_{c}$. Here, we will neglect
this extra splitting and consider an effective ``heavy'' degenerate
$d_{xz/yz}$ electron-like band $1$, whose bottom is located $\mu_{c}+\mu$
below the Fermi level, and a ``light'' electron-like $d_{xy}$ band
$2$, whose bottom is located $2\mu_{c}$ above the bottom of the
band 1 (see Fig. \ref{2band}(a)). Notice that in the LAO/STO interfaces
(shown in Fig. \ref{2band}(b)), there is an orbital reconstruction
that shifts the $d_{xy}$ band below the $d_{xz/yz}$ \cite{sall:09}.
While this will change the possible orbital occupations of the electrons,
the effective two-band model considered here is still representative
of the system.

Introducing the electronic creation operators $c_{a,\mathbf{k}\sigma}^{\dagger}$,
with band index $a$, momentum $\mathbf{k}$, and spin $\sigma$,
we have the following interacting Hamiltonian in the pairing (particle-particle)
channel:

\begin{eqnarray}
H & = & \sum_{a,\mathbf{k},\sigma}\left(\varepsilon_{a,\mathbf{k}}-\mu\right)c_{a,\mathbf{k}\sigma}^{\dagger}c_{a,\mathbf{k}\sigma}+\label{H}\\
 &  & \sum_{a,b,\mathbf{k},\mathbf{k}'}V_{ab,\mathbf{k}\mathbf{k}'}c_{a,\mathbf{k}\uparrow}^{\dagger}c_{a,-\mathbf{k}\downarrow}^{\dagger}c_{b,-\mathbf{k}'\downarrow}c_{b,\mathbf{k}'\uparrow}+\mathrm{h.c.}\nonumber 
\end{eqnarray}
 where $\varepsilon_{a,\mathbf{k}}$ gives the band dispersion $\varepsilon_{1,k}=k^{2}/2m_{1}-\mu_{c},\varepsilon_{2,k}=k^{2}/2m_{2}+\mu_{c}$,
$\mu$ is the chemical potential, and $V_{ab,\mathbf{k}\mathbf{k}'}$
are the pairing interactions. Hereafter, we assume them to be momentum-independent,
i.e. we consider s-wave SC states and for convenience choose the zero
of energy to be at the midpoint between the two energy band minima.
Furthermore, we consider that the pairing involves only states near
the Fermi level, yielding the mean-field $2\times2$ system of gap
equations:

\begin{equation}
\Delta_{i}=-\sum_{j=1}^{2}\rho_{j}V_{ij}\Delta_{j}\int_{\mu_{j}}^{W}d\xi\,\frac{\tanh\left(\frac{\sqrt{\xi^{2}+\Delta_{j}^{2}}}{2T}\right)}{2\sqrt{\xi^{2}+\Delta_{j}^{2}}}\label{gap_equations}
\end{equation}

Here, $W$ is the upper cutoff of the interaction, $\Delta_{i}$ are
the SC gaps, $\rho_{i}$ are the density of states at the Fermi level
(in 2D the DOS is assumed to be independent of energy), $\mu_{1}=-\mu-\mu_{c}$
is the position of the bottom of the electron-band 1, and $\mu_{2}=\mu_{c}-\mu$
is the position of the bottom of the second band. For simplicity,
we will discuss our results in terms of the dimensionless coupling
constants $\lambda_{ij}=-\rho_{i}V_{ij}$. We assume $V_{21}=V_{12}$
and therefore $\lambda_{21}=(\rho_{2}/\rho_{1})\lambda_{12}$. 

From Eqs. (\ref{gap_equations}) it is straightforward to obtain $T_{c}$.
One linearizes the gap equations and calculates their largest eigenvalue,
which equals $1$ at $T_{c}$. This procedure yields the implicit
equation:

\begin{multline}
1=\left(\frac{\lambda_{11}I_{1}\left(\mu\right)+\lambda_{22}I_{2}\left(\mu\right)}{2}\right)\label{eqs_Tc}\\
+\sqrt{\left(\frac{\lambda_{11}I_{1}\left(\mu\right)-\lambda_{22}I_{2}\left(\mu\right)}{2}\right)^{2}+\kappa\lambda_{12}^{2}I_{1}\left(\mu\right)I_{2}\left(\mu\right)}
\end{multline}
where $\kappa\equiv\rho_{2}/\rho_{1}$ is the ratio between the density
of states of the two bands at the Fermi level and $I_{a}\left(\mu\right)$
are implicit functions of $T_{c}\equiv1.13W\,\mathrm{e}^{-1/\lambda_{c}}$:

\begin{eqnarray}
I_{1}\left(\mu\right) & = & \frac{1}{\lambda_{c}}+\frac{1}{2}\ln\left(\frac{\mu+\mu_{c}}{W}\right)\label{aux_Tc}\\
I_{2}\left(\mu\right) & = & \frac{\theta\left(\mu\right)}{2}\left[\frac{1}{\lambda_{c}}+\mathrm{sign}\left(\mu-\mu_{c}\right)\int_{0}^{\frac{\left|\mu-\mu_{c}\right|}{2T_{c}}}dx\,\frac{\tanh x}{x}\right]\nonumber 
\end{eqnarray}

We note that $T_{c}\ll\mu_{c}\ll W$, which justifies the use of the
log functions above. The last term in $I_{2}\left(\mu\right)$ is
the one responsible for the changes in $T_{c}$ as the second band
crosses the Fermi level. In particular, very close to the point $\mu=\mu_{c}$,
this function has a linear dependence on $\mu-\mu_{c}$, whereas far
from this point, it behaves as the logarithm of $\mu-\mu_{c}$. We
can also obtain the gap functions $\Delta_{i}$ from Eqs. (\ref{gap_equations})
in a straightforward way. For $T=0$, we obtain simple algebraic equations
for $\Delta_{i0}$. For non-zero $T<T_{c}$, one has to solve the
self-consistent gap equations numerically.

\begin{figure}
\includegraphics[width=1\columnwidth]{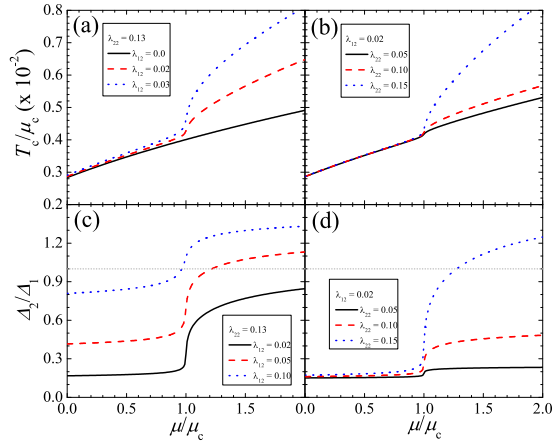} \caption{(Color Online) (a-b) Superconducting transition temperature $T_{c}$
and (c-d) the $T$ = 0 ratio of the gap as a function of chemical
potential for various values of $\lambda_{12}$ and $\lambda_{22}$.
The light gray dotted line refers to $\Delta_{2}$/$\Delta_{1}$ =
1. Note, the gap of the second band is finite at all doping, but it
is significantly smaller for $\mu<\mu_{c}$. In all panels, $\lambda_{11}=0.14$.}

\label{Tc} 
\end{figure}

Multi-band gap equations such as (\ref{gap_equations}) have been
investigated in the past in the context of different compounds, such
as iron pnictides and MgB$_{2}$. Here, our interest is an aspect
of these gap equations that has not been as widely investigated \cite{Fernandes_Schmalian,Bianconi},
namely, the evolution of the solutions of the gap equations as a function
of the chemical potential - particularly near the point where the
second band starts to be occupied, $\mu=\mu_{c}$.

\subsection{Results for $T_{c}$ and $\Delta_{i}$}

We first need to discuss the values of the parameters that describe
the electron-doped STO compounds. First, we note that the two bands
have different orbital content, i.e. band $1$ is composed of the
$d_{xz/yz}$ orbitals, whereas band $2$ is composed of $d_{xy}$
orbitals. Since intra-orbital interactions are typically larger than
inter-orbital interactions, we expect the intra-band pairing to be
larger than the inter-band pairing, i.e. $\left|\lambda_{ii}\right|\gg\left|\lambda_{i\neq j}\right|$.
As we will show below, this gives results in agreement with the experimental
observations of Binnig $et~al.$

Under these conditions, SC only appears for attractive intra-band
interactions, $\lambda_{ii}>0$, regardless of the sign of $\lambda_{12}$,
since only $\lambda_{12}^{2}$ appears in the equation (\ref{eqs_Tc})
for $T_{c}$. The sign of the inter-band coupling $\lambda_{12}$,
however, affects the structure of the SC state, as the eigenvector
that diagonalizes the linearized form of Eqs. (\ref{gap_equations})
is such that $\mathrm{sign}\left(\Delta_{1}\Delta_{2}\right)=\mathrm{sign}\left(\lambda_{12}\right)$.
Thus, for $\lambda_{12}>0$ (attractive inter-band interaction $V_{12}<0$),
$\Delta_{1}$ and $\Delta_{2}$ have the same sign (the so-called
$s^{++}$ state), whereas for $\lambda_{12}<0$ (repulsive inter-band
interaction $V_{12}>0$), $\Delta_{1}$ and $\Delta_{2}$ have opposite
signs (the $s^{+-}$ state) \cite{Fernandes_Schmalian}. Since the
SC free energy depends only on $\lambda_{12}\Delta_{1}\Delta_{2}$,
the thermodynamic properties discussed in this paper do not depend
on the relative sign between $\Delta_{1}$ and $\Delta_{2}$. Therefore,
we set $\lambda_{12}>0$, but it is possible that STO is an unconventional
s-wave superconductor with $s^{+-}$ structure.

In Fig. \ref{Tc}, we present the enhancement of $T_{c}$ as well
as the ratio between the zero-temperature gaps $\Delta_{20}/\Delta_{10}$
as a function of $\mu/\mu_{c}$ for several different values of $\lambda_{12}$
and $\lambda_{22}$, with $\lambda_{11}=0.14$ fixed. Since band $2$
is lighter than band $1$, we have $\rho_{2}=(m_{2}/m_{1})\rho_{1}=0.4\rho_{1}$,
in accordance with band structure calculations \cite{Mattheiss72}.
For the upper-cutoff $W$, we set $W=10\mu_{c}$, and we consider
$\lambda_{11}\approx\lambda_{22}\ll1$ as well as $\lambda_{12}\ll\lambda_{ii}$,
as discussed above. As it is shown in the figure, even when the inter-band
pairing is small, the occupation of the second band at $\mu=\mu_{c}$
is marked by a sharp increase in $T_{c}$ and by an enhancement in
the zero-temperature gaps $\Delta_{10}$ and $\Delta_{20}$. While
for $\mu<\mu_{c}$ the gap $\Delta_{10}$ is significantly larger
than $\Delta_{20}$, for $\mu>\mu_{c}$ they become of similar magnitudes.
The complete temperature behavior of the gap functions is shown in
Fig. \ref{Egap-3DLDOS}(a), which corresponds roughly to two BCS-like
curves.

If we considered a larger inter-band interaction $\lambda_{12}$,
then we would have obtained a larger enhancement of $T_{c}$ at $\mu=\mu_{c}$,
but the gap function $\Delta_{20}$ would be much closer to $\Delta_{10}$
even for $\mu<\mu_{c}$ (shown in Fig. \ref{Tc}(a) and (c)). Meanwhile,
the tunneling experiments of Binnig \emph{et al. }\cite{Bednorz80}
suggest instead that $\Delta_{20}$ is considerably smaller than $\Delta_{10}$
near $\mu=\mu_{c}$. Conversely, if we considered a smaller intra-band
interaction $\lambda_{22}\ll\lambda_{11}$, then the enhancement of
$T_{c}$ at $\mu=\mu_{c}$ would be rather weak, in disagreement with
the same data. Therefore, unless some fine-tuning is claimed, our
results put important constraints on the character of the multi-band
SC of electron-doped STO, as it appears to be dominated by attractive
intra-band pairing interactions and weaker inter-band pairing.

Note that this situation is the opposite of other multi-band superconductors,
such as the iron pnictides. These materials have a much larger repulsive
inter-band interaction $\lambda_{12}^{2}\gg\lambda_{11}\lambda_{22}$
enhanced by spin fluctuations, which gives rise to a sign-reversal
$s^{+-}$ state \cite{reviews_FePn,Fernandes_Schmalian}. Contrary
to the iron pnictides, in STO the SC state is not near other electronic
ordered states whose fluctuations can enhance $\lambda_{12}$ .

\section{Tunneling spectrum \label{sec_tunneling}}

The existence of two SC gaps has important implications for several
thermodynamic quantities. For instance, two-gap superconductivity
can be inferred from the temperature dependence of the specific heat
and penetration depth. Yet, the most direct probe for multi-gap superconductivity
is the local density of states (LDOS), which can be measured via tunneling
spectroscopy.

\begin{figure}
\includegraphics[width=3.5in]{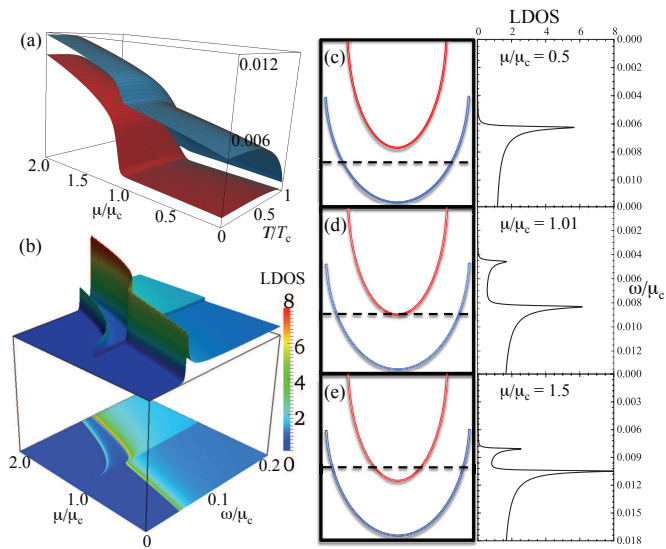} \caption{(Color Online) (a) Superconducting gap as a function of chemical potential
and temperature, displaying the standard BCS temperature dependence.
(b) Calculated LDOS for the two-band model as function of increasing
chemical potential and energy. (c)-(e) The chemical potential levels
with respect to the two bands (left) and the calculated LDOS for the
two-band model for $\mu$/$\mu_{c}$ = (c) 0.5, (d) 1.01, and (e)
1.5 and for the set of parameters $\lambda_{11}=0.14$, $\lambda_{22}=0.13$,
and $\lambda_{12}=0.02$}

\label{Egap-3DLDOS} 
\end{figure}

Within our two-band model, the total LDOS $N\left(\omega\right)$
is the sum of the contributions from each band:

\begin{equation}
N(\omega)=-\frac{1}{\pi}\,{\rm Im}\sum_{{\bf k}}[\nu_{1}G_{11}(\omega,{\bf k})+\nu_{2}G_{22}(\omega,{\bf k})]\label{aux_LDOS}
\end{equation}
where $\nu_{1,2}$ denotes the degeneracy of the band. Within the
mean field approximation each Green's function takes the BCS form
$G_{jj}=(\omega+\xi_{j}({\bf k}))/[\omega^{2}-\xi_{j}^{2}({\bf k})-\Delta_{j}^{2}]$
with the respective gaps from Eq. \ref{gap_equations}. By replacing
the summation over momenta by an integration over energies, the LDOS
can be calculated analytically, yielding

\begin{equation}
N(\omega)={\rm Re}\left[\sum_{j}\frac{\left(\omega+i\eta+c_{j}\right)\rho_{j}(c_{j})+\left(\omega-c_{j}\right)\rho_{j}(-c_{j})}{2c_{j}}\right]\label{LDOS}
\end{equation}

\noindent where $\rho_{j}$ are the density of states of each band,
$c_{j}$ = $\sqrt{(\omega+i\eta)-\Delta_{j}^{2}}$ , and $\eta$ provides
a small width to simulate experimental measurements. For the two-dimensional
model considered so far, $\rho_{j}$ are constants.

Using the $T=0$ gaps obtained from the solution of Eqs. (\ref{gap_equations}),
we calculate $N\left(\omega\right)$. We observe two different behaviors,
as shown in Fig. \ref{Egap-3DLDOS}(b). For $\mu<\mu_{c}$, $N\left(\omega\right)$
has a sharp peak near $\omega$ = $\Delta_{10}$, and the smaller
gap $\Delta_{20}$ shows no noticeable signature. This is due to the
fact that this gap is not open at the Fermi level, but at an energy
near the bottom of the second band $\mu_{c}-\mu\gg\Delta_{20}$. We
note that the LDOS has a clear increase at $\omega=\mu_{c}-\mu$ due
to the additional contribution of the states coming from the second
band. Only when $\mu_{c}-\mu\sim\Delta_{20}$, we see a weak signature
of the second gap at energies slightly larger than $\Delta_{10}$,
where a clear peak is still observed.

For $\mu>\mu_{c}$, after the second band crosses the Fermi level,
the second gap gives rise to a smaller peak at energies lower than
the first peak, since $\Delta_{20}<\Delta_{10}$. As $\mu$ increases,
this peak becomes larger and moves closer to the peak coming from
the first-band gap. Eventually these two peaks become difficult to
distinguish, since the gaps become comparable in magnitude. As a result,
the LDOS has a sharp peak preceded by a shoulder-like feature (shown
in Fig. \ref{Egap-3DLDOS}(c)-(e)). These general features of the
LDOS spectrum are in good agreement with the tunneling data of Ref.
\cite{Bednorz80}.

\section{three-dimensional case \label{sec_three_dimensional} }

\begin{figure}
\includegraphics[width=0.95\columnwidth]{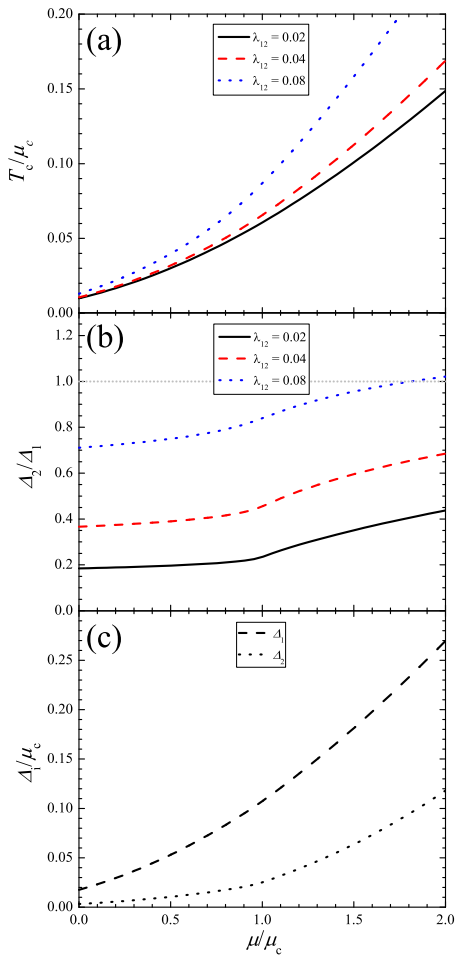} \caption{(Color Online) Calculations for the 3D two-band model as function
of chemical potential for $\lambda_{11}$ = 0.14 and $\lambda_{22}$
= 0.13. We present $T_{c}$ (a) and $\Delta_{2}$/$\Delta_{1}$ (b)
for $\lambda_{12}$ = 0.02, 0.04, and 0.08. The light gray dotted
line in (b) across $\Delta_{2}$/$\Delta_{1}$ = 1 denotes the point
at which the second gap crosses the first gap. The individual energy
gaps ($\Delta_{i}$) for $\lambda_{12}$ = 0.02 are shown in (c).}

\label{3D-2band} 
\end{figure}

So far we have focused on the 2D case, which is most relevant to STO
thin films, interfaces, or probes sensitive to surface states, such
as the tunneling experiments by Binnig \emph{et al.} \cite{Bednorz80}.
In the bulk STO material, the 3D dispersion gives rise to a density
of states $\rho\left(\xi\right)\propto\sqrt{\xi}$, which is expected
to smoothen the sharp features observed in $T_{c}$ and $\Delta_{i0}$
at $\mu=\mu_{c}$. Nevertheless, as we will show in this section,
the enhancement of $T_{c}$ and $\Delta_{i0}$ at $\mu=\mu_{c}$ is
still present. 

We continue to assume parabolic bands. Then, the density of states
of each band is given by $\rho_{i}=\rho_{i,0}\sqrt{\epsilon+\mu\pm\mu_{c}}$
and the gap equations are changed accordingly to:

\begin{eqnarray}
 &  & {\displaystyle {\displaystyle \Delta_{i}}=-\sum_{j=1}^{2}\rho_{j,0}V_{ij}\Delta_{j}\times}\label{3D_gap_eqs}\\
 &  & {\displaystyle \int_{\mu_{j}}^{W}d\xi\,\frac{\sqrt{\epsilon+\mu\pm\mu_{c}}}{2\sqrt{\xi^{2}+\Delta_{j}^{2}}}\tanh\left(\frac{\sqrt{\xi^{2}+\Delta_{j}^{2}}}{2T}\right)}\nonumber 
\end{eqnarray}

The solution of this system of equations is straightforward. In Fig.
\ref{3D-2band}, we show the detailed behavior of the 3D two-band
model as function of the chemical potential for $\lambda_{11}=0.14$
and $\lambda_{22}=0.13$ (the same parameters as in the 2D case).
The evolution of $T_{c}$ does produce an ``upturn\textquotedbl{}
around $\mu=\mu_{c}$, as shown in Fig. \ref{3D-2band}(a). The ratio
of the gaps, presented in Fig. \ref{3D-2band}(b), displays a kink
as the second band is populated. However, as shown in Fig. \ref{3D-2band}(c),
the individual energy gaps display a distinct increase. This is consistent
with the data observed by Binnig \emph{et al} \cite{Bednorz80}, where
$T_{c}$ and the gaps clearly increase as function of the chemical
potential. The $T_{c}$ data in Binnig \emph{et al} \cite{Bednorz80}
shows a slight kink as the second band becomes populated. The 3D model
does not explain this slight kink and may indicate the presence of
a 2D band structure from surface states. Therefore, we conclude that
a two dimensional model is more accurate to describe the STO systems
and may be the leading cause for the creation of a 2D electron gas
at the interface of LAO and STO.

\begin{figure}
\includegraphics[width=0.9\columnwidth]{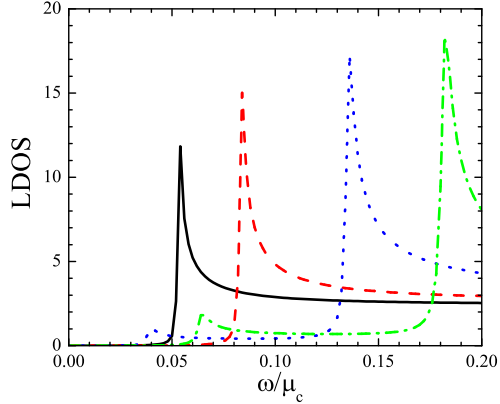} \caption{(Color Online) Calculated LDOS for the weakly coupled 3D two-band
model as a function of $\omega/\mu_{c}$ for $\mu$/$\mu_{c}$ = 0.5
(solid-black), 0.8 (dashed-red), 1.2 (dotted-blue), 1.5 (dash-dotted-green).
Here, we use the parameters $\lambda_{11}$ = 0.14, $\lambda_{22}$
= 0.13, $\lambda_{12}$ = 0.02 and $\kappa$ = 0.4.}

\label{3D-LDOS} 
\end{figure}

For completeness, we also calculate the local density of states (LDOS)
resulting from the three-dimensional, two-band model. The LDOS is
also given by Eq. (\ref{LDOS}), but with the appropriate energy-dependent
density of states $\rho_{i}$. Figure \ref{3D-LDOS} shows the results,
with the LDOS plotted as a function of $\omega/\mu_{c}$ for $\mu/\mu_{c}$
= 0.5, 0.8, 1.2, 1.5. The behavior is similar to the one observed
by Binnig \emph{et al.} in Ref. \cite{Bednorz80}, where the population
of the second band produces a small lower-frequency shoulder and an
increase in the LDOS.

\section{conclusions \label{sec_conclusions}}

The observation of multiple gap structures in the tunneling spectrum
of electron-doped bulk STO \cite{Bednorz80} is a well known fact,
yet it has been largely overlooked in the discussion on the origin
of superconductivity in both doped STO and LAO/STO intefaces. Motivated
by this observation we proposed a multi-band model for superconductivity
in electron-doped STO. The d-band manifold is split into two $d_{xy}$
and $d_{xz/yz}$ bands with a splitting of around $10$ meV before
doping. We show that with increasing electronic occupation, the higher
d-band(s) become occupied, engaging an additional source of electrons
in the pairing. The onset of the occupation of the higher bands results
in an increased DOS and pairing condensate gain, which in turn results
in a sharp increase in $T_{c}$ and in the SC gap in the second band.
Comparing to experiments, we find the primary source of pairing to
be the intra-band coupling $\lambda_{11,22}$ with smaller inter-band
Josephson coupling $\lambda_{12}\ll\lambda_{11,22}$. Depending on
the sign of the inter-band coupling, the SC gaps on the two bands
can have the same or opposite signs, resulting in either $s^{++}$
or $s^{+-}$ SC states. Although our results were motived by bulk
STO, the fact that $T_{c}$ in the LAO/STO interfaces is very close
to the $T_{c}$ of electron-doped STO \cite{koon:67}, and the observation
that the maximum $T_{c}$ of the heterostructure takes place when
a second d-band becomes occupied \cite{huij:09,gari:09}, we suggest
that the same multi-band effects may take place in the oxide interfaces.
Of course, in the heterostructures, the breaking of inversion symmetry
at the surface may lead to additional contributions to $T_{c}$, not
considered in our model \cite{Haraldsen12}. To test our proposal,
we suggest to use simple direct spectroscopy in the SC state of doped
STO thin films and LAO/STO interfaces, such as scanning tunneling
microscopy or planar tunneling. Both probes would reveal the existence
of a two gap structure if indeed superconductivity is a two-band phenomena
in these systems. The observation of two band features would likely
be possible only in a clean limit when scattering will not wash away
features of the weaker gap.

\medskip{}

We are grateful to Q. Jia, P. Hirschfeld, H. Hwang, T. Kopp, J. Mannhart,
I. Schuller, J. Triscone, S. Trugman, J.X. Zhu for useful discussions
and criticism. This work was supported by US BES E304, LDRD and Aspen
Center for Physics. Los Alamos National Laboratory, an affirmative
action equal opportunity employer, is operated by Los Alamos National
Security, LLC, for the National Nuclear Security Administration of
the U.S. Department of Energy under contract DE-AC52-06NA25396. R.M.F.
and A.V.B. were supported by the NSF Partnerships for International
Research and Education (PIRE) program OISE-0968226.

\end{document}